\documentclass{ws-ijmpe}
\usepackage{graphics}
\usepackage{epsfig}
\usepackage{color}
\usepackage{pstricks}
\def\lsim{\buildrel < \over {_{\sim}}}
\def\gsim{\buildrel > \over {_{\sim}}}
\def\Lmunu{L_{\mu\nu}}
\def\Wmunu{W^{\mu\nu}}
\def\openone{\leavevmode\hbox{\small1\kern-3.8pt\normalsize1}}
\newcommand{\beq}{\begin{equation}}
\newcommand{\eeq}{\end{equation}}
\newcommand{\be}{\begin{eqnarray}}
\newcommand{\ee}{\end{eqnarray}}
\newcommand{\qt}{{\widetilde q}}
\begin{document}

\markboth{Omar Benhar}{Many-Body Theory of the 
Electroweak Nuclear Response}

\catchline{}{}{}{}{}

\title{MANY-BODY THEORY OF THE \\
ELECTROWEAK NUCLEAR RESPONSE
}

\author{\footnotesize Omar Benhar
}

\address{INFN and Department of Physics, Universit\`a ``La Sapienza''\\
I-00185 Roma, Italy \\
benhar@roma1.infn.it}

\maketitle

\begin{history}
\received{(received date)}
\revised{(revised date)}
\end{history}

\begin{abstract}
After a brief review of the theoretical description of nuclei 
based on nonrelativistic many-body theory and realistic hamiltonians, these
lectures focus on its application to the analysis of the electroweak response. 
Special emphasis is given to electron-nucleus scattering, whose experimental 
study has provided a wealth of information on nuclear structure and dynamics, 
exposing the limitations of the shell model. 
The extension of the formalism to the case of neutrino-nucleus interactions, 
whose quantitative understanding is required to reduce the 
systematic uncertainty of neutrino oscillation experiments, is also 
discussed.
\end{abstract}

\section{Introduction}

Over the past four decades, electron scattering has provided a
wealth of information on nuclear structure and dynamics.
Form factors and charge distributions have been extracted from
elastic scattering data, while inelastic measurements have allowed
for a systematic study of the dynamic response over a broad range
of momentum and energy transfer.\cite{BDS}
In addition, with the advent of the last generation of continuous beam 
accelerators, a number of exclusive processes
have been analyzed with unprecedented precision (for recent reviews, see, 
e.g. Ref.\cite{eepreviews}). 

In electron scattering experiments the nucleus il mostly seen
as a target. Studying its interactions with the probe,
whose properties are completely specified, one obtains information 
on the unknown features of its internal structure. 
                                                                                           
The emerging picture clearly shows that the nuclear shell model fails to 
provide a fully quantitative description of the existing data. 
More realistic many body approaches, in which correlation 
effetcs are explicitely taken into account, appear to be needed 
to explain electron scattering observables. 
                                                                                           
In neutrino oscillation experiments, on the other hand, nuclear interactions 
are exploited to {\em detect} the beam particles, whose kinematics is largely unknown.
Using the nucleus as a detector obviously requires that its response
to neutrino interactions be under control. Fulfillment of this prerequisite 
is in fact critical to keep the systematic uncertainty associated with the
reconstruction of the neutrino kinematics to an acceptable level.\cite{nuint}

These lectures are aimed at providing an introduction to 
nuclear many-body theory and its application to the calculation of the
electron- and neutrino-nucleus cross section. Section \ref{NMBT} contains
a short overview of the underlying dynamical model and formalism. 
For pedagogical purposes, the nuclear response is first analyzed
in the case of a scalar probe in Section \ref{SQW}, while the generalization 
to electron and 
charged current neutrino interactions is discussed in Sections \ref{eA} and
\ref{nuA}, respectively. 
Finally, Section \ref{summa} is devoted to a summary of the main results.

\section{Non relativistic Nuclear Many-Body Theory}
\label{NMBT}

In nuclear many-body theory (NMBT) the nucleus is viewed as a collection of
$A$ pointlike protons and neutrons, whose dynamics are described by the
nonrelativistic hamiltonian
\beq
H = \sum_i \frac{{\bf p}_i^2}{2m} + \sum_{j>i} v_{ij}
+ \sum_{k>j>i} V_{ijk}\ ,
\label{hamiltonian}
\eeq
where ${\bf p}_i$ and $m$ denote the momentum of the $i$-th nucleon and
the nucleon mass, respectively.
The two body potential $v_{ij}$ is determined by fitting deuteron properties
and $\sim$ 4000 precisely measured nucleon-nucleon (NN) scattering phase shitfs.\cite{WSS}
It turns out to be strongly spin-isospin dependent and non central, and reduces to the 
Yukawa one pion exchange potential at large separation distance. 
The inclusion of the three-nucleon interaction, satisfying 
$\langle~V_{ijk}~\rangle~\ll~\langle v_{ij}~\rangle$,
is required to account for the binding energy of the three-nucleon systems.\cite{PPCPW}

The many body Schr\"odinger equation associated with the hamiltonian
of Eq.(\ref{hamiltonian}) can be solved exactly, using stochastic methods,
for nuclei with mass number $A\leq 12$. The resulting energies
of the ground and low-lying excited states are in excellent agreement
with experimental data.\cite{WP}

It has to be emphasized that the dynamics of NMBT is fully determined by 
the properties of exactly solvable system, and does not suffer from the 
uncertainties involved in many-body calculations, which unavoidably make use of
approximations. Once the nuclear hamiltonian is determined, calculations
of the properties of a variety of nuclear systems, ranging from deuteron to
neutron stars, can be carried out {\em without making use of any adjustable
parameters}.

The main difficulty associated with the use of the hamiltonian of Eq.(\ref{hamiltonian}) 
in a many-body calculation lies in the strong repulsive
core of the NN force, which cannot be handled within standard
perturbation theory. 

In the shell model, this problem is circumvented replacing the interaction
terms in Eq.(\ref{hamiltonian}) with a {\em mean field}, according to
\beq
\sum_{j>i} v_{ij} + \sum_{k>j>i} V_{ijk} \longrightarrow \sum_{i} U_{i} \ .
\eeq
Within this scheme, the many-body Schr\"odinger equation reduces to a
trivial single particle problem, and the ground state wave function can
be written in the form
\beq
\Phi_0(1,\ldots,A) = {\mathcal A}
\prod_{\alpha_i \epsilon \{ F \}} \phi_{\alpha_i}(i) \ ,
\label{sm:wf}
\eeq
where the antisymetrization operator takes into account Pauli principle and
the $\phi_{\alpha_i}(i)$ are solutions of the eigenvalue equations
\beq
\left( \frac{ {\bf p}_i^2 }{ 2 m } + U_{i} \right) \phi_{\alpha_i}(i) =
\epsilon_{\alpha_i} \phi_{\alpha_i}(i) \ .
\eeq
The product appearing in Eq.(\ref{sm:wf}) includes the states belonging
to the Fermi sea $\{F\}$, i.e. the $A$ lowest energy states.

In the Fermi gas (FG) model, in which neglecting interactions are neglected 
altogether, Eq.(\ref{sm:wf})
describes a degenerate gas of nucleons occupying all momentum eigenstates
belonging to the eigenvalues $|{\bf k}| < k_F$. The Fermi momentum $k_F$
is related to the nucleon density $\rho = A/V$, $V$ being the normalization volume,
through $k_F = 2 \rho/3 \pi^2$.

In spite of its simplicity, the shell model provides a remarkably good
description of a number of nuclear properties.
However, the results of electron- and hadron-induced nucleon knock-out
experiments have provided overwhelming evidence of its inadequacy to account for the
full complexity of nuclear dynamics.

While the spectroscopic lines corresponding to knock-out from shell model orbits can
be clearly identified in the measured energy spectra, the corresponding strengths
turn out to be consistently and sizably lower than expected, regardless of the
nuclear mass number.

This discrepancy is mainly to be ascribed to the effect of dynamical correlations
induced by the NN forces, whose effect in not taken into account in the shell
model. Correlations give rise to virtual scattering processes, leading to the
excitation of the participating nucleons to states of energy larger than the
Fermi energy, thus depleting the shell model states within the Fermi sea.

The first realistic treatment of the nuclear many-body problem was based on 
G-matrix perturbation theory, developed by Br\"uckner, Bethe and Goldstone.\cite{DayRev} 
Within this approach the nuclear hamiltonian is rewritten in the form (we neglect the three-body
potential, for simplicity)
\beq
H = H_0 + H_I 
\eeq
with
\beq
H_0 = \sum_i \left( \frac{{\bf p}_i^2}{2m} +W_i \right) \ \ , \ \ 
H_I = \sum_{j>i} v_{ij} - \sum_i W_i \ .
\eeq
The single particle potential $W_i$ is chosen in such a way as to either
simplify the calculations or satisfy analitycity properties.
The problem of the divergences arising from the short range repulsion of the 
bare interaction is circumvented replacing the NN potential $v$ with 
the $G$-matrix, obtained summing up {\em ladder} diagrams at all orders in 
the perturbative expansion through the integral equation
\beq
G=v-v\frac{Q}{e}\ G,      
\label{reaction:matrix}
\eeq
where $Q$ is a projection operator accounting for the effects of Pauli blocking 
and $e$ is an energy denominator. 

A widely employed alternative approach exploits the possibility of 
embodying non perturbative 
effects 
in the basis functions. 
This is the foundation of Correlated Basis Function (CBF) perturbation 
theory,\cite{Fee69,CMK79,FPa88} which provides a consistent and unified treatment of 
light nuclei and nuclear matter. 

The {\em correlated} states are obtained from the eigenstates of the shell model 
hamiltonian through the transformation
\beq
|n \rangle = \frac{ F| n ) }{ ( n | F^{\dag} F | n )^{1/2}} \ ,  
\label{basis}
\eeq
where 
$F$ is a correlation operator, whose 
structure reflects the properties of the NN potential. It is written in the form
\beq
F=\mathcal{S} \ \prod_{j>i} f_{ij} \ ,
\eeq
where $\mathcal{S}$ is the symmetryzation operator and
\beq
f_{ij} =\sum_n f^n(r_{ij})O^n_{ij} \ .   
\label{F:operator}
\eeq
The minimal set of operators $O^n_{ij}$ includes the four central components
associated with the different spin-isospin channels ($n=1,4$) and the isoscalar and 
isovector tensor components ($n=5,6$): 
\beq
O^p_{ij} = [1, (\bm{\sigma}_{i}\cdot\bm{\sigma}_{j}), S_{ij}]
\otimes[1,(\bm{\tau}_{i}\cdot\bm{\tau}_{j})] ~,
\eeq
with 
\beq
S_{ij} = \frac{3}{r^2}({\bm \sigma}_i \cdot {\bf r})
({\bm \sigma}_j \cdot {\bf r}) - ({\bm \sigma}_i \cdot {\bm \sigma}_j)\ .
\eeq
In some calculation the spin orbit (${\bf L}\cdot{\bf S}$) and isovector spin 
orbit ($({\bf L}\cdot{\bf S})({\bm\tau}_i\cdot{\bm\tau}_j)$) components, 
needed to reproduce the NN scattering phase shifts in $S$ and $P$ wave, have
been also included.

The radial functions $f^n(r_{ij})$ are determined through functional minimization of
the expectation value 
\beq
\langle H \rangle = \frac{\langle 0 | H | 0 \rangle}{\langle 0 | 0 \rangle} \ ,
\eeq
evaluated using the cluster expansion formalism.\cite{CMK79} 

The correlated wave functions $|n\rangle$ 
are required to exhibit two important properties:\cite{Fee69} the 
{\em cluster factorization} 
property, dictated by the finite range of the interaction, and the {\em core}
property, due to the short range repulsion. The former property implies 
that $f^{n=1}(r) \rightarrow 1$ and $f^{n>1}(r) \rightarrow 0$ as $r \rightarrow \infty$, 
while the core property requires that all correlation functions become vanishingly 
small at short interparticle distance. 

The basis of correlated states, while being complete, is {\em not} orthogonal. However,
it can be orthogonalized using standard techniques of many-body theory.\cite{FPa88}
As the operator $F^\dagger H F$ is well behaved, as is the $G$-matrix, the perturbative
expansions of nuclear observable in the correlated basis are rapidly convergent.
 
\section{The nuclear response to a scalar probe}
\label{SQW}

Within NMBT, the nuclear response to a scalar probe delivering
momentum {\bf q} and energy $\omega$ can be written in terms of the the 
imaginary part of the polarization propagator $\Pi({\bf q},\omega)$ according 
to\cite{FetterWalecka,BFF:1992}
\beq
\label{def:resp}
S({\bf q},\omega) = \frac{1}{\pi}\ {\rm Im}\  \Pi({\bf q},\omega) = \frac{1}{\pi}\
{\rm Im}\ \langle 0 \vert \rho^\dagger_{{\bf q}}  \
\frac{1}{H-E_0-\omega-i\eta} \ \rho_{{\bf q}} \vert 0 \rangle \ ,
\eeq
where $\eta=0^+$, $\rho_{{\bf q}}= \sum_{{\bf k}} a^\dagger_{{\bf k}+{\bf q}} a_{{\bf k}}$
is the operator describing the fluctuation of the target density induced by the interaction
with the probe,
$a^\dagger_{{\bf k}}$ and $a_{{\bf k}}$ are nucleon creation
and annihilation operators, and $\vert 0 \rangle$ is the
target ground state, satisfying the Schr\"odinger equation
$H\vert 0 \rangle = E_0 \vert 0 \rangle$.

In this Section, we will discuss the relation between $S({\bf q},\omega)$ and 
the nucleon Green function, leading to the popular expression of the response in 
terms of spectral functions.\cite{BFF:1992,BFF:1989} This discussion
is mainly aimed at showing that the spectral function formalism, while 
being often advocated using heuristic arguments, can be derived in a rigorous 
and fully consistent fashion.
For the sake of simplicity, we will consider uniform nuclear matter with
equal numbers of protons and neutrons. 

Equation (\ref{def:resp}) clearly shows that the interaction with the probe leads to
a transition of the struck nucleon from a {\em hole state} of momentum ${\bf k}$, 
with $|{\bf k}|<k_F$, to a {\em particle state} of momentum ${\bf k}+{\bf q}$, 
with $|{\bf k}+{\bf q}|>k_F$.
The calculation of $S({\bf q},\omega)$ amounts to describing the propagation of 
the resulting particle-hole pair through the nuclear medium.

The fundamental quantity involved in the theoretical treatment of many-body systems is
the Green function, i.e. the quantum mechanical amplitude associated with the propagation
of a particle from $x\equiv(t,{\bf x})$ to $x^\prime\equiv(t^\prime,{\bf x}^\prime)$
.\cite{FetterWalecka} In uniform matter, due to translation invariance,
the Green function only depends on the difference $x-x^\prime$, and after Fourier
transformation to the conjugate variable $k~\equiv~(~{\bf k}~,~E~)$ can be 
written in the form
\be
\nonumber
G({\bf k},E) & = & \langle 0 \vert a^\dagger_{{\bf k}} \
\frac{1}{H-E_0-E-i\eta} \ a_{{\bf k}} \vert 0 \rangle
 - \langle 0 \vert a_{{\bf k}} \ \frac{1}{H-E_0+E-i\eta} \ a^\dagger_{{\bf k}} \vert 0 \rangle \\ & = &  G_h({\bf k},E) + G_p({\bf k},E) \ ,
\label{green:1}
\ee
where $G_h$ and $G_p$ correspond to propagation of nucleons in hole and
particle states, respectively.

The connection between Green function and spectral functions is established
through the Lehman representation\cite{FetterWalecka}
\beq
G({\bf k},E) = \int dE^\prime \left[ \frac{P_h({\bf k},E^\prime)}{E^\prime - E - i\eta}
    -  \frac{P_p({\bf k},E^\prime)}{E - E^\prime - i\eta} \right] \ ,
\eeq
implying
\be
P_h({\bf k},E) & = & \sum_n
\vert \langle n_{(N-1)}(-{\bf k}) \vert a_{{\bf k}} \vert 0_N \rangle \vert^2
\delta(E-E^{(-)}_n+E_0) = \frac{1}{\pi}\ {\rm Im}\ G_h({\bf k},E) \ , \\
\label{def:Ph}
P_p({\bf k},E) & = & \sum_n
\vert \langle n_{(N+1)}({\bf k}) \vert a^\dagger_{{\bf k}} \vert 0_N \rangle \vert^2
\delta(E+E^{(+)}_n-E_0) = \frac{1}{\pi} {\rm Im}\ G_p({\bf p},E) \ ,
\label{def:Pp}
\ee
where $\vert \langle n_{(N \pm 1)}(\pm {\bf k}) \rangle$ denotes an eigenstate of the
$(A \pm 1)$-nucleon system, carrying momentum $\pm{\bf k}$ and energy $E^{(\pm)}_n$.

Within the FG model the matrix elements of the creation and annihilation operators
reduce to step functions, and the Green function takes a very simple form.
For example, for hole states we find\footnote{Note that, according to our 
definitions, the hole spectral function is defined for $E \geq -\mu$, $\mu$ being 
the Fermi energy.}
\beq
G_{FG,h}({\bf k},E) = \frac{ \theta(k_F-|{\bf k}|) }{ E+\epsilon^0_k-i\eta }  \ ,
\label{green:FG}
\eeq
with $\epsilon^0_k = |{\bf k}^2| /2m$, implying
\beq
P_{FG,h}({\bf k},E) = \theta(k_F-|{\bf k}|) \delta(E+\epsilon^0_k) \ .
\eeq
Strong interactions modify the energy of a nucleon carrying momentum ${\bf k}$ according to
$\epsilon^0_k \longrightarrow \epsilon^0_k + \Sigma({\bf k},E)$, where $\Sigma({\bf k},E)$
is the {\em complex} nucleon self-energy, describing the effect of nuclear dynamics.
As a consequence, the Green function for hole states becomes
\beq
G_h({\bf k},E) = \frac{1}{ E + \epsilon^0_k + \Sigma({\bf k},E)} \ .
\label{greenh:2}
\eeq

A very convenient decomposition of $G_h({\bf k},E)$
can be obtained inserting a complete set of $(A-1)$-nucleon states
(see Eqs.(\ref{green:1})-(\ref{def:Ph})) and isolating the contributions of
one-hole {\em bound} states, whose weight is given by\cite{BFF:1990}
\beq
Z_k = | \langle -{\bf k} | a_{{\bf k}} | 0 \rangle |^2 = \theta(k_F-|{\bf k}|)
\Phi_k \ .
\label{def:Z}
\eeq
Note that in the FG model these are the only nonvanishing terms,
and $\Phi_k \equiv 1$, while in the presence of interactions
$\Phi_k < 1$.
The resulting contribution to the Green function exhibits a pole at
$-\epsilon_k$, the {\em quasiparticle} energy $\epsilon_k$ being defined
by the equation
\beq
\epsilon_k = \epsilon^0_k + {\rm Re}\ \Sigma({\bf k},\epsilon_k) \ .
\label{QP:energy}
\eeq
The full Green function can be rewritten
\beq
G_h({\bf k},E) = \frac{Z_k}{E+\epsilon_k+i Z_k\ {\rm Im}\ \Sigma({\bf k},e_k)}
 + G^B_h({\bf k},E) \ ,
\eeq
where $G^B_h$ is a smooth contribution, asociated with $(A-1)$-nucleon states
having at least one nucleon excited to the continuum (two hole-one particle,
three hole-two particles \ldots) due to virtual scattering processes
induced by nucleon-nucleon (NN) interactions. The corresponding spectral function is
\beq
P_h({\bf k},E) =  \frac{1}{\pi}\
\frac{ Z_k^2 \ {\rm Im}\ \Sigma({\bf k},\epsilon_k) }
{ [E + \epsilon^0_k + {\rm Re}\ \Sigma({\bf k},\epsilon_k)]^2 +
          [Z_k {\rm Im}\ \Sigma({\bf k},\epsilon_k)]^2 }
 + P^B_h({\bf k},E) \ .
\eeq
The first term in the right hand side of the above equation yields the spectrum of
a system of independent quasiparticles, carrying momenta $|{\bf k}|<k_F$, moving in
a complex mean field whose real and imaginary parts determine the quasiparticle
effective mass and lifetime, respectively. The presence of the second term is
a consequence of nucleon-nucleon correlations, not taken into account in the mean
field picture. Being the only one surviving at $|{\bf k}|>k_F$, in the FG model
this correlation term vanishes.

\begin{figure}[th]
\centerline{\psfig{file=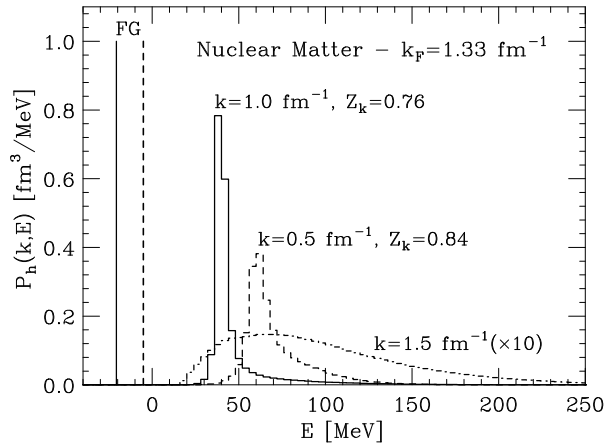,width=8cm}}
\vspace*{8pt}
\caption{ Energy dependence of the hole spectral function of nuclear 
matter.\protect\cite{BFF:1989} The solid, dashed and dot-dash lines 
correspond to $|{\bf k}|=$ 1, 0.5 and 1.5 fm$^{-1}$,
respectively. The FG spectral function at $|{\bf k}|=$ 1 and 0.5 fm$^{-1}$ is shown
for comparison.
The quasiparticle strengths of Eq.(\ref{def:Z}), are also reported. \label{fig1} }
\end{figure}

Figure \ref{fig1} illustrates the energy dependence of the hole spectral function
of nuclear matter, calculated in Ref.\cite{BFF:1989} using CBF perturbation theory and
a realistic nuclear hamiltonian. Comparison with the FG model clearly shows
that the effects of nuclear dynamics and NN correlations are large, resulting
in a shift of the quasiparticle peaks, whose finite width becomes
large for deeply-bound states with $|{\bf k}| \ll k_F$. In addition, NN
correlations are responsible
for the appearance of strength at $|{\bf k}|>k_F$. The energy integral
\beq
\label{def:nk}
n(k) = \int dE\  P_h({\bf k},E)
\eeq
yields the occupation probability of the state of momentum ${\bf k}$. The results
of Fig. \ref{fig1} clearly show that in presence of correlations
$n(|{\bf k}|>k_F)\neq0$.

In general, the calculation of the response requires the knowledge of $P_h$ and $P_p$, as
well as of the particle-hole effective interaction.\cite{BFF:1992,WIM:2004}
The spectral functions are mostly affected by short range NN correlations
(see Fig. \ref{fig1}), while the inclusion of the effective interaction, e.g. within the
framework of the Random Phase Approximation (RPA),\cite{WIM:2004} is needed to account
for collective excitations induced by long range correlations, involving more than
two nucleons.

At large momentum transfer, as the space resolution of the probe becomes small compared
to the average NN separation distance, $S({\bf q},\omega)$ is no longer significantly
affected by long range correlations. In this kinematical regime the zero-th order
approximation in the effective interaction, is expected to be applicable. Whithin this 
scenario, the response reduces to the incoherent sum of contributions coming from
scattering processes involving a single nucleon, and can be written in the simple form
\beq
\label{L0}
S({\bf q},\omega) = \int d^3k dE\ P_h({\bf k},E) P_p({\bf k}+{\bf q},\omega-E) \ .
\eeq
The widely employed impulse approximation (IA) can be readily obtained from
the above definition replacing $P_p$ with the FG result, which amounts to
disregarding final state interactions (FSI) betwen the struck nucleon and the spectator
particles. The resulting expression reads
\beq
\label{IA}
S_{IA}({\bf q},\omega) = \int d^3k dE\ P_h({\bf k},E) \theta(|{\bf k}+{\bf q}|-k_F)
\delta(\omega-E-\epsilon^0_{|{\bf k}+{\bf q}|}) \ .
\eeq

\begin{figure}[th]
\centerline{\psfig{file=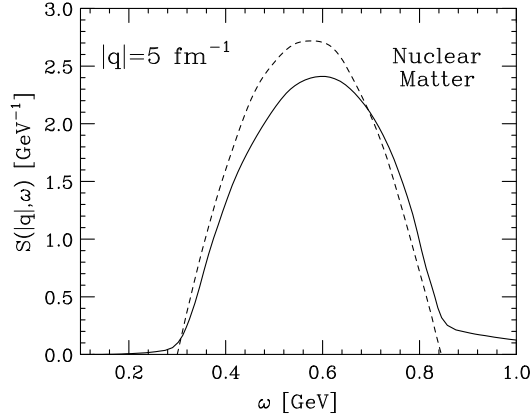,width=7cm}}
\vspace*{8pt}
\caption{ Nuclear matter $S_{IA}({\bf q},\omega)$ (see Eq.(\ref{IA})), as a function
of $\omega$ at $|{\bf q}|=5$ fm$^{-1}$.  The solid and dashed lines correspond to
the spectral function of Ref.\protect\cite{BFF:1989} and to the FG model (shifted in such
a way as to account for nuclear matter binding energy), respsctively. \label{iscorr} }
\end{figure}

Figure \ref{iscorr}, showing the $\omega$ dependence of the nuclear matter structure 
function at $|{\bf q}|=5$ fm$^{-1}$, illustrates the role of correlations 
in the target ground state. The solid and dashed lines have been obtained from 
Eq.(\ref{IA}) using the spectral function of Ref.\cite{BFF:1989} and that resulting
from the FG model (shifted in such a way as to account for nuclear matter binding energy), 
respectively. It clearly appears that the inclusion of correlations produces a 
significant shift of the strength towards larger values of energy transfer.

At moderate momentum transfer, both the full response and the particle and hole
spectral functions can be obtained using non relativistic many-body theory.
The results of Ref.\cite{BFF:1992} suggest that the zero-th order approximations
of Eqs.(\ref{L0}) and (\ref{IA}) are fairly accurate at $|{\bf q}|~\gsim~500$~MeV.
However, in this kinematical
regime the motion of the struck nucleon in the final state can no longer
be described using the non relativistic formalism.
While at IA level this problem can be easily circumvented replacing the
non relativistic kinetic energy with its relativistic counterpart,
including the effects of FSI in the response of Eq.(\ref{L0})
involves further approximations, needed to obtain the particle spectral function
at large $|{\bf q}|$.

A systematic scheme to include corrections to Eq.(\ref{IA}) and take into
account FSI, originally proposed in Ref.\cite{gangofsix}, is discussed
in Ref.\cite{marianthi}. The main effects of FSI on the response are i) a 
shift in energy, due to the mean field of the spectator nucleons and ii) a 
redistributions of the strength, due to the coupling of the one particle-one hole 
final state to $n$ particle-$n$ hole final states.

In the simplest implementation of the approach of Refs.\cite{gangofsix,marianthi}, 
the response 
is obtained from the IA result according to
\beq
S({\bf q},\omega) = \int d\omega^\prime\ S_{IA}({\bf q},\omega^\prime)
 f_{{\bf q}}(\omega-\omega^\prime) \ ,
\label{S:fold}
\eeq
the folding function $f_{{\bf q}}$ being related to the particle spectral function
through
\beq
P_p({\bf k}+{\bf q},\omega-E) = \theta(k_F-|{\bf k}+{\bf q}|) \
 f_{|{\bf k}+{\bf q}|}(\omega-E-e^0_{|{\bf k}+{\bf q}|})
\label{f:fold}
\eeq
with $\epsilon^0_{|{\bf k}+{\bf q}|} = \sqrt{|{\bf k}+{\bf q}|^2+m^2}$.
In the absence of FSI, $f_{{\bf q}}$ shrinks to
a $\delta$-function and the IA result of Eq.(\ref{IA}) is recovered.

Obvioulsy, at large ${\bf q}$ the calculation of $P_p({\bf k}+{\bf q},\omega-E)$ cannot
be carried out using a nuclear potential model. Hovever, it can be
obtained form the measured NN scattering amplitude within the eikonal
approximation. The resulting folding function is
the Fourier transform of the Green function describing the 
propagation of the struck particle, travelling in the direction of the $z$-axis 
with constant velocity $v$:
\beq
f_{|{\bf q}|}(\omega) = \int \frac{dt}{2 \pi} \ {\rm e}^{i\omega t} 
  {\rm e}^{i\int_0^t dt^\prime {\widetilde V}_{|{\bf q}|}(vt^\prime)} \ . 
\eeq
where ${\bf k}+{\bf q} \approx {\bf q}$ and 
\beq
{\widetilde V}_{|{\bf q}|}(z) =  \langle 0 |  \frac{1}{A}
\sum_{j>i} \Gamma_{|{\bf q}|}({\bf r}_{ij} + {\bf z}) | 0  \rangle \ .
\label{aver:V}
\eeq
In the above equation, $\Gamma_{|{\bf q}|}$ is the Fourier transform
of the NN scattering amplitude at incident mometum $|{\bf q}|$ and momentum 
transfer $|{\bf t}|$, $A_{|{\bf q}|}(k)$, parametrized according to
\beq
A_{|{\bf q}|}(p) = 
\frac{|{\bf q}|}{4 \pi} \sigma (i + \alpha){\rm e}^{-\beta p^2} \ .
\eeq
In principle, the total cross section $\sigma$, the slope $\beta$ and the
ratio between the real and the imaginary part, $\alpha$, can be extracted
from NN scattering data. However, the modifications of the scattering amplitude
due to the presence of the nuclear medium are known to be sizable, and must 
be taken into account. The calculation of these corrections within the 
framework of NMBT is discussed in Ref.\cite{papi}.

In Eq.(\ref{aver:V}), the expectation value is evaluated in the {\em correlated}
ground state. It turns out that NN correlation, whose
effect on $P_h$ is illustrated in Fig. \ref{fig1}, also
affect the particle spectral function and, as a consequence, the folding function
of Eq. (\ref{f:fold}). Neglecting all correlations
\beq
{\widetilde V}_{|{\bf q}|}(z) \rightarrow {\widetilde V}^0_{|{\bf q}|}
 = \frac{1}{2} v \rho \sigma(i + \alpha) \ ,
\eeq 
and the quasiparticle approximation 
\beq
P_p({\bf q},\omega-E) =
\frac{1}{\pi} \frac{ {\rm Im}\  {\widetilde V}^0_{|{\bf q}|} }
{ \left[ \omega-E-e^0_{|{\bf q}|}-{\rm Re}\ {\widetilde V}^0_{|{\bf q}|} \right]^2
 + \left[ {\rm Im}\ {\widetilde V}^0_{|{\bf q}|} \right]^2 }
\eeq
is recovered. 

\begin{figure}[th]
\centerline{\psfig{file=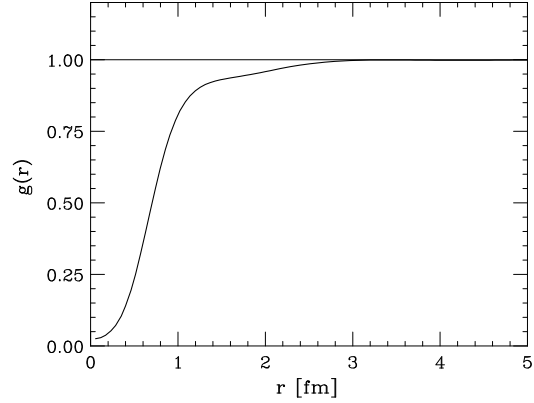,width=7cm}}
\vspace*{8pt}
\caption{Radial distribution function of nuclear matter at equilibrium 
density, obtained from CBF perturbation theory using a realistic 
hamiltonian. \label{gofr} }
\end{figure}

Correlations induce strong density fluctuations, preventing two nucleon from 
coming close to one another. The joint probability of finding two particles 
at positions ${\bf r_1}$ and ${\bf r_2}$ can be written 
\beq
\rho({\bf r_1},{\bf r_2}) = \rho^2 g(|{\bf r_1}-{\bf r_2}|)
\eeq
where the $g(r)$ is the {\em radial distribution function}, shown in Fig.\ref{gofr}. 

The effect of correlation on FSI can be easily understood: as the probability of 
finding a spectator within the range of the repulsive core of 
the NN force ($\lsim 1$ fm) is small, the probability that the struck particle 
rescatter against one of the spectators within a length $\sim 1/|{\bf q}|$ 
(the space resolution of the probe) is also very small at large $|{\bf q}|$. 
Hence, inclusion of correlations leads to a significant suppression of FSI 
effects.\cite{gangofsix,marianthi}

\begin{figure}[th]
\centerline{\psfig{file=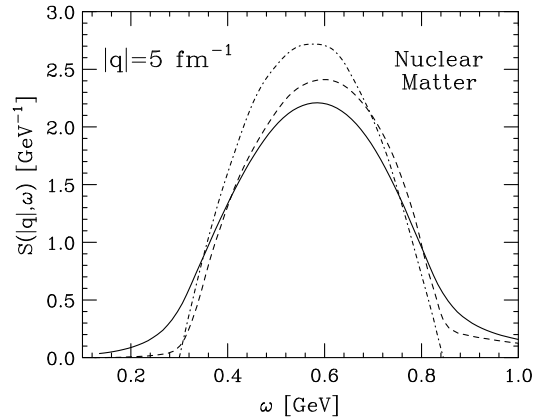,width=7cm}}
\vspace*{8pt}
\caption{Nuclear matter $S({\bf q},\omega)$, defined in Eq.(\ref{S:fold}), as a function
of $\omega$ at $|{\bf q}|=5$ fm$^{-1}$.  The solid and dashed lines have been obtained
from the spectral function of Ref.\protect\cite{BFF:1989}, with and without inclusion 
of FSI, respectively. The dot-dash line corresponds to the FG model (shifted in such
a way as to account for nuclear matter binding energy), respectively. \label{fsicorr} }
\end{figure}

Fig. \ref{fsicorr} shows the $\omega$ dependence of the nuclear matter response of 
Eq.(\ref{S:fold}) at $|{\bf q}|=5$ fm$^{-1}$. The solid and dashed lines have been 
obtained using the spectral function of Ref.\cite{BFF:1989}, with and without inclusion
of FSI according to the formalism of Ref.\cite{gangofsix}, respectively.
For reference, the results of the FG model are also shown by the dot-dash line.
The two effects of FSI, energy shift and redistribution of the strength from the region of 
the peak to the tails, clearly show up in the comparison betweem soild and dashed lines.

\section{The electron-nucleus cross section}
\label{eA}

The differential cross section of the process
\beq
e + A \rightarrow e^\prime + X \ ,
\label{eA:process}
\eeq
in which an electron of initial four-momentum $k_e\equiv(E_e,{\bf k}_e)$ scatters off
a nuclear target to a state of four-momentum
$k^\prime_e\equiv(E_{e^\prime},{\bf k}_{e^\prime})$, the target final state 
being undetected, can be written in Born approximation as\cite{Itzykson80}
\beq
\frac{d^2\sigma}{d\Omega_{e^\prime} dE_{e^\prime}} =
\frac{\alpha^2}{Q^4}\frac{E_{e^\prime}}{E_e}\ \Lmunu\Wmunu \ ,
\label{eA:xsec}
\eeq
where $\alpha=1/137$ is the fine structure constant,
$d\Omega_{e^\prime}$ is the differential solid angle in the direction
specified by ${\bf k}_{e^\prime}$, $Q^2=-q^2$ and
$q=k_e-k_{e^\prime} \equiv (\omega,{\bf q})$ is the four momentum transfer.

The tensor $\Lmunu$, that can be written neglecting the lepton mass as
\beq
\Lmunu = 2 \left[ k_e^\mu k_{e^\prime}^\nu + k_e^\nu k_{e^\prime}^\mu
 - g^{\mu\nu} (k_e k_{e^\prime}) \right] \ ,
\label{lepten}
\eeq
where $g^{\mu \nu} \equiv(1,-1,-1,-1)$ and $(k_e k_{e^\prime})=E_e
E_{e^\prime}-{\bf k}_e \cdot {\bf k}_{e^\prime}$, is fully specified by the measured
electron kinematical
variables. All the information on target structure is contained
in the tensor $\Wmunu$, whose definition involves the initial and final nuclear
states $|0\rangle$ and $|X\rangle$, carrying four-momenta $p_0$ and $p_X$,
as well as the nuclear current operator $J^\mu$:
\beq
\Wmunu=\sum_{X}  \langle 0 |J^\mu|X \rangle\langle X|J^\nu|0\rangle
\delta^{(4)}(p_0+q-p_X)\ ,
\label{nuclear:tensor}
\eeq
\noindent
where the sum includes all hadronic final states. Note that the tensor 
of Eq.(\ref{nuclear:tensor}) is the generalization of the nuclear response, 
discussed in Section \ref{SQW}, to the case of a probe interacting 
with the target through a vector current. This can be easily seen inserting 
\beq
\sum_n | n \rangle \langle n | = 1	\ ,
\eeq
$|n\rangle$ being an eigenstate of the nuclear hamiltonian sasisfying 
$H|n\rangle=E_n|n\rangle$, in the definition of Eq.(\ref{def:resp}). The result is
\beq
S({\bf q},\omega) = \sum_n  \langle 0 | \rho^\dagger_{{\bf q}} | n \rangle 
\langle n | \rho_{{\bf q}} | 0 \rangle \delta(\omega + E_0 - E_n ) \ ,
\eeq
to be compared to Eq.(\ref{nuclear:tensor}).

The most general expression of the target tensor of Eq.~(\ref{nuclear:tensor}),
fulfilling the
requirements of Lorentz covariance, conservation of parity and gauge invariance, can be
written in terms of two structure functions $W_1$ and $W_2$ as
\beq
\Wmunu = W_1 \left( -g^{\mu\nu} + \frac{q^\mu q^\nu}{q^2} \right) 
 + \frac{W_2}{M^2} \left(p_0^\mu - \frac{(p_0 q)}{q^2}q^\mu \right)
                   \left(p_0^\nu - \frac{(p_0 q)}{q^2}q^\nu \right) \ ,
\label{genw}
\eeq
where $M$ is the target mass and the structure functions depend on the two scalars $Q^2$
and $(p_0 q)$. In the target rest frame $(p_0 q) = m\omega$ and $W_1$ and $W_2$ become
functions of the measured momentum and energy transfer $|{\bf q}|$ and ${\omega}$.
        
Substitution of Eq.~(\ref{genw}) into Eq.~(\ref{eA:xsec}) leads to
\beq
\frac{d^2\sigma}{d\Omega_{e^\prime} dE_{e^\prime}} = 
\left( \frac{d\sigma}{d\Omega_{e^\prime}}\right)_M 
 \left[ W_2(|{\bf q}|,\omega)
            + 2 W_1(|{\bf q}|,\omega) \tan^2\frac{\theta}{2} \right] \ ,
\label{eA:xsec12}
\eeq
where $\theta$ and
$(d\sigma/d\Omega_{e^\prime})_M= \alpha^2 \cos^2(\theta/2)/4E_e^2\sin^4(\theta/2)$
denote the electron scattering angle and the Mott cross section, respectively.

At moderate momentum transfer, typically  $( {\bf |q|}~< 0.5\,~{\rm GeV})$, the 
response tensor $\Wmunu$ of Eq.~(\ref{nuclear:tensor}) can be obtained from 
NMBT, using nonrelativistic wave functions to describe the initial and final
states and expanding the current operator in powers of ${\bf |q|}/m$.

Exact calculations of the electron-nucleus cross section can be carried out 
for light nuclei, with $A\leq4$, either 
solving the Schr\"odinger equation for bound and continuum states\cite{Golak95}
or using integral transform techniques.\cite{Efros94,Carlson98} 
The latter approach is discussed in detail in Prof. Leidemann's lectures.\cite{Leidemann}
Accurate calculations are also possible for uniform nuclear matter, 
as translation invariance considerably simplifies the problem.\cite{FP,FF}
On the othe hand, the available results for medium-heavy targets have been 
mostly obtained using the mean field approach, supplemented by the inclusion of
model residual interactions to take into account long range 
correlations.\cite{Dellafiore85} 

At higher values of ${\bf |q|}$, corresponding to beam energies larger than $\sim$1 GeV,
 describing the final states $|X\rangle$ in terms of nonrelativistic nucleons is 
no longer possible. Due to the prohibitive difficulties involved in a fully consistent 
treatment of the relativistic nuclear many-body problem, calculations of $\Wmunu$ in 
this regime require a set of simplifying assumptions, allowing one to take into account 
the relativistic motion of final state particles carrying momenta $\sim {\bf q}$, as 
well as inelastic processes leading to the production  of hadrons other than protons 
and neutrons. 

\subsection{The impulse approximation}

\begin{figure}[th]
\centerline{\psfig{file=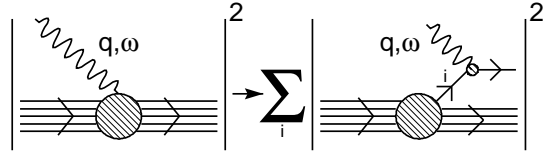,width=8cm}}
\vspace*{8pt}
\caption{Schematic representation of the IA regime, in which
the nuclear cross section is replaced by the incoherent sum of
cross sections describing scattering off individual nucleons, the
recoiling $({\rm A}-1)$-nucleon system acting as a spectator.\label{cartoon}} 
\end{figure}

Within the IA picture, schematically represented in Fig. \ref{cartoon}, the nuclear 
current appearing in Eq.~(\ref{nuclear:tensor}) is written as a sum of one-body 
currents
\beq
J^\mu \rightarrow \sum_i j_i^\mu \ ,
\label{currIA}
\eeq
while $|X\rangle$ reduces to the direct product of the hadronic state produced at the
electromagnetic vertex, carrying momentum ${\bf p}_x$, and the state describing the
residual system, carrying momentum
${\bf p}_{\cal R}= {\bf q}-{\bf p}_x$ (in order to simplify the notation, spin
indices will be omitted)
\beq
|X\rangle \rightarrow |x,{\bf p}_x\rangle
\otimes |{\cal R},{\bf p_{\cal R}}\rangle \ .
\label{resIA}
\eeq

Using Eq.~(\ref{resIA}) we can replace
\beq
\sum_X | X \rangle \langle X |  \rightarrow \sum_{x}
\int d^3p_x  | x,{\bf p}_x \rangle \langle {\bf p}_x,x | 
 \sum_{{\cal R}} d^3p_{{\cal R}}
| {\cal R}, {\bf p}_{{\cal R}} \rangle \langle {\bf p}_{{\cal R}}, {\cal R} | \ .
\label{sumn}
\eeq
Substitution of Eqs.~(\ref{currIA})-(\ref{sumn}) into Eq.~(\ref{nuclear:tensor}) and
insertion of a complete set of free nucleon states leads to the factorization of the 
nuclear current matrix element. As a result, the {\em incoherent} contribution to 
Eq.~(\ref{nuclear:tensor}) can be 
rewritten in the form
\beq
\Wmunu({\bf q},\omega) = \int d^3k \ dE \
\left(\frac{m}{E_{{\bf k}}}\right) \left[ Z P_p({\bf k}, E)
w_p^{\mu\nu}({\widetilde q}) +  N P_n({\bf k}, E) 
w_n^{\mu\nu}({\widetilde q}) \right] \ ,
\label{hadrten2}
\eeq
where $Z$ and $N=A-Z$ are the number of target protons and neutrons, while 
$P_p$ and $P_n$ denote the proton and neutron {\em hole} spectral functions, 
respectively. In Eq.~(\ref{hadrten2}), $E_{{\bf k}} = \sqrt{|{\bf k}^2|+m^2}$ and
\beq
w_N^{\mu\nu} = \sum_x \langle {\bf k},{\rm N}| j^\mu_N | x,{\bf k}+{\bf q} \rangle
\langle {\bf k}+{\bf q},x | j^\nu_N | {\rm N},{\bf k} \rangle  
 \delta({\widetilde \omega} + E_{{\bf k}} - E_x ) \  ,
\label{nucleon:tensor}
\eeq
with\cite{BDS} 
\beq
{\widetilde \omega} = E_x - E_{{\bf k}} = E_0 + \omega - E_{{\cal R}}
- E_{{\bf k}} = \omega - E + m - E_{{\bf k}} \ .
\label{omega:tilde}
\eeq
The above equations show that within the IA scheme, the definition of the
electron-nucleus cross section involves two elements: i) the tensor
$w_N^{\mu\nu}$, defined by Eq.~(\ref{nucleon:tensor}), describing the electromagnetic
interactions of a {\it bound} nucleon carrying momentum ${\bf k}$ and ii) the spectral
function, discussed in the previous Sections.

\subsection{Electron scattering off a bound nucleon}
\label{sec:eNxsec}

While in electron-nucleon scattering in free space the struck particle is given
the entire four momentum transfer $q\equiv(\omega,{\bf q})$, in a scattering process
involving a bound nucleon a fraction $\delta \omega$ of the energy loss goes
into the spectator system. This mechanism emerges in a most natural fashion from
the IA formalism.

Assuming that the current operators are not modified by the nuclear environment,
the quantity defined by Eq.~(\ref{nucleon:tensor}) can be identified with
the tensor describing electron scattering off a {\it free} nucleon at four
momentum transfer $q\equiv({\bf q},{\widetilde \omega})$. Hence, Eq.~(\ref{nucleon:tensor})
shows that within IA binding is taken into account through the replacement
\beq
q\equiv(\omega,{\bf q}) \rightarrow {\widetilde q}\equiv({\widetilde \omega},
{\bf q}) \ .
\label{qtilde}
\eeq
The interpretation of $\delta \omega = \omega - {\widetilde \omega}$ as the amount of
energy going into the recoiling spectator system becomes particularly transparent
in the limit $|{\bf k}|/m \ll 1$, in which Eq.~(\ref{omega:tilde}) yields
$\delta \omega = E$.

The tensor $w_N^{\mu\nu}$ of Eq.~(\ref{nucleon:tensor}) can be obtained from the 
general expression (compare to Eq.~(\ref{genw}))
\beq
w_N^{\mu\nu}  = w^N_1 \left( -g^{\mu\nu} + \frac{\qt^\mu \qt^\nu}{\qt^2} \right) 
 + \frac{w^N_2}{m^2} \left(k^\mu - \frac{(k \qt)}{\qt^2}\qt^\mu \right)
                   \left(k^\nu - \frac{(k \qt)}{\qt^2}\qt^\nu \right) \ ,
\label{genw2}
\eeq
where $k\equiv (E_{\bf k},{\bf k})$ and the two structure functions $w_1$ and $w_2$ can 
be extracted from the measured electron-proton and electron-deuteron scattering 
cross sections\cite{BR,bm1}. 

For example, in the case of quasielastic scattering 
 $w_1$ and $w_2$ are
simply related to the electric and magnetic nucleon form factors,
$G_{E_N}$ and $G_{M_N}$, through
\beq
w^N_1 = -\frac{\qt^2}{4m^2}\ \delta\left({\widetilde \omega} + \frac{\qt^2}{2m} \right)
 \ G_{M_N}^2 \ ,
\label{qe1}
\eeq
\beq
w^N_2 = \frac{1}{1 - \qt^2/4 m^2} \ \delta\left({\widetilde \omega} +
\frac{\qt^2}{2m} \right) 
 \left( G_{E_N}^2 - \frac{\qt^2}{4m^2} G_{M_N}^2 \right) \ .
\label{qe2}
\eeq
A similar expression can be used to
describe resonance production. The explicit formula differ in the analytical form
of the relevant form factors and for the replacement of the energy conserving
$\delta$-function with a Breit-Wigner factor, accounting for the finite width of the
resonance.\cite{bm2}
Finally, Eq.(\ref{genw2}) can be applied in the region of deep inelastic scattering,  
where the structure functions $F^N_1 = m w^N_1$ and $F^N_2 = \omega w^N_2$ 
depend on $Q^2=|{\bf q}^2|- \omega^2$ and the bjorken scaling 
variable $x = Q^2/2m\omega$.

As a final remark, it has to be pointed out that 
the replacement of $\omega$ with $\tilde{\omega}$, while being reasonable on physics
grounds, and in fact quite natural in the context of the IA analysis, poses a
considerable conceptual problem, in that it leads to a violation of current
conservation, that requires
\beq
q_\mu w_N^{\mu\nu} = 0 \ .
\label{gauge:inv}
\eeq
However, violation of gauge invariance in the IA scheme 
turns out to be only marginally relevant to
inclusive electron scattering at large momentum transfer. The results of 
numerical studies suggest that the main effect of nuclear binding can indeed be 
accounted for with the replacement $\omega \rightarrow {\widetilde \omega}$.\cite{BDS}

\subsection{Comparison to electron scattering data}

The formalism outlined above, supplemented by the treatment of FSI effects
proposed in Refs.\cite{gangofsix,marianthi}, has been widely and successfully
applied to the analysis of electron-nucleus scattering data.\cite{BDS}

In Ref. \cite{Benhar05}, it has been employed
to calculate the inclusive electron scattering cross sections off oxygen at beam
energies ranging between 700 and 1200 MeV and electron scattering angle 32$^\circ$.
In this kinematical region, relevant to many neutrino experiments, single nucleon
knock out is the dominant reaction mechanism and both quasi-elastic
and inelastic processes, leading to the appearance of nucleon resonaces,
must be taken into account.

Comparison between theoretical results and the experimental data of
Ref. \cite{LNF} shows
 that, while the data in the region of the quasi-elastic peak are accounted for
with an accuracy better than $\sim$ 10 \%, theory fails to explain the measured cross
sections at larger electron energy loss, where $\Delta$ production
dominates.

\begin{figure}[hbt]
{\psfig{figure=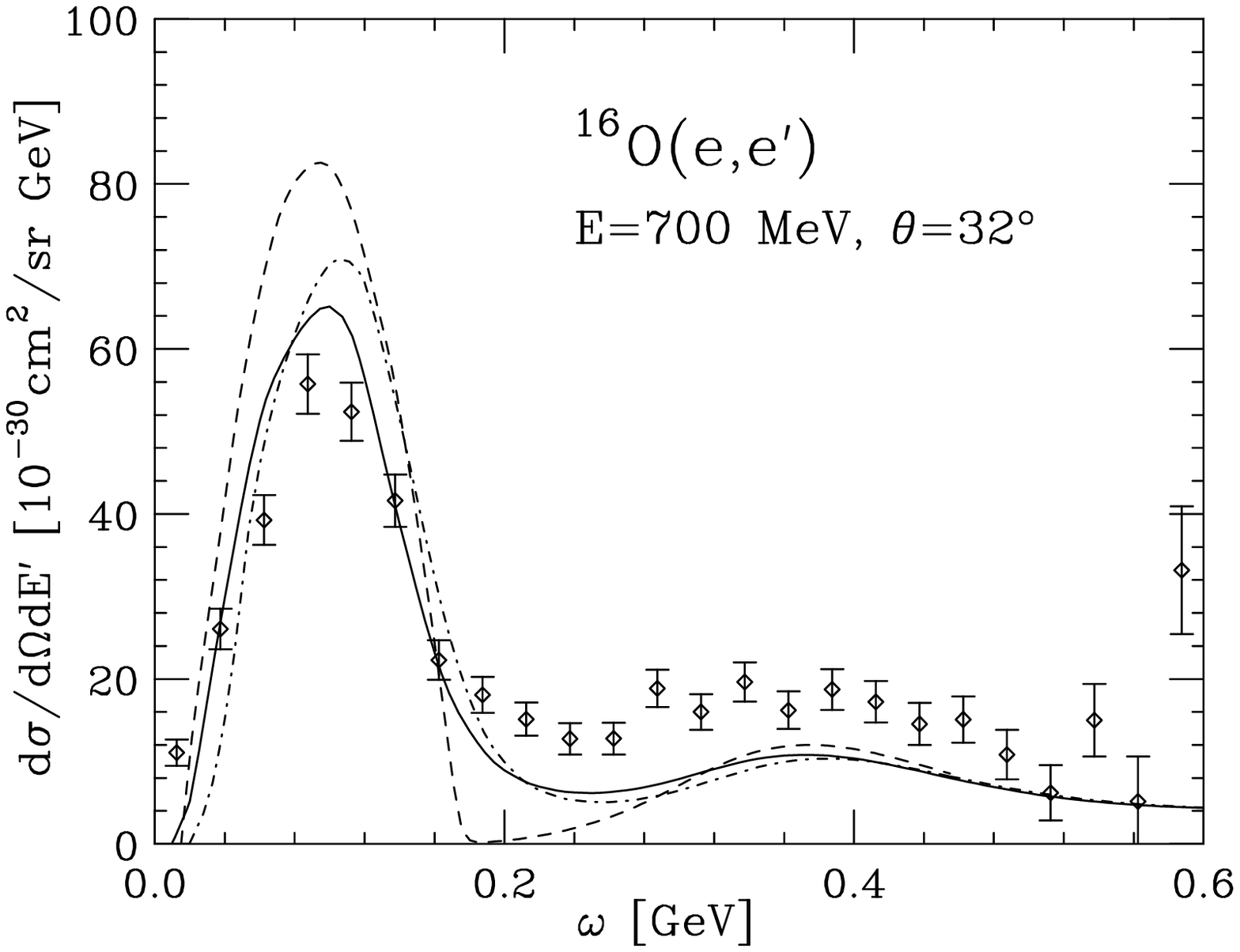,angle=00,width=6.2cm,height=5.0cm}}
{\psfig{figure=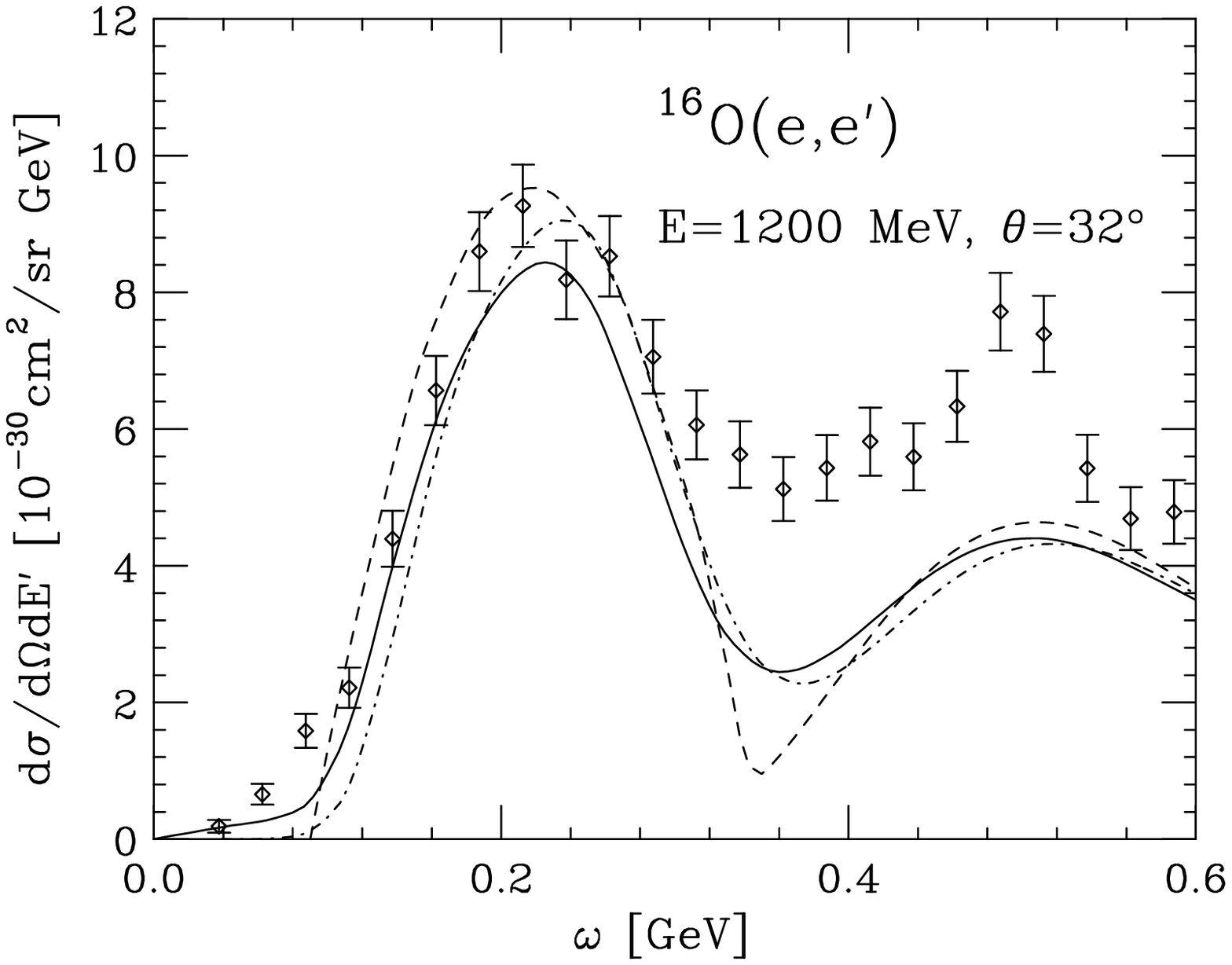,angle=00,width=6.2cm,height=5.0cm}}
\vspace*{8pt}
\caption{ Cross section of the process $^{16}O(e,e^\prime)$
at scattering angle 32$^\circ$ and beam energy 700 MeV (left panel) and 1200 MeV
(right panel), as a function of the electron energy loss $\omega$. 
Solid lines: full calculation, including FSI. Dot-dash lines: IA calculation.
Dashed lines: FG model with $p_F = 225$ MeV and $\epsilon = 25$ MeV. The data are taken 
from Ref.\protect\cite{LNF} \label{fig:ee}}
\end{figure}

As an example, Fig. \ref{fig:ee} shows the results of Ref.\cite{Benhar05} at
beam energy 700 and 1200 MeV. For reference, the results of the Fermi gas (FG) model
corresponding to Fermi momentum $p_F = 225$ MeV and average removal energy
$\epsilon = 25$ MeV are also shown. Theoretical calculations have been carried
out using the spectral function of Ref.\cite{BFFS}, the H\"ohler-Brash parameterization 
of the nucleon form factors \cite{Hohler76,Brash02} in the quasi-elastic channel and the
Bodek and Ritchie parametrization of the proton and neutron structure functions
in the inelastic channels \cite{BR}.

The authors of Ref. \cite{Benhar05} argued that the disagreement
between theory and data in the $\Delta$ production region is likely to be imputable
to deficiencies in the description of the neutron structure 
functions at low $Q^2$.\footnote{In the kinematics of Fig. \ref{fig:ee} 
the $\Delta$ production peak corresponds to 
$Q^2 \sim 0.2$ GeV$^2$.} This conclusion is supported by the analysis 
carried out in.\cite{bm1,hiroki}

\begin{figure}[th]
{\psfig{figure=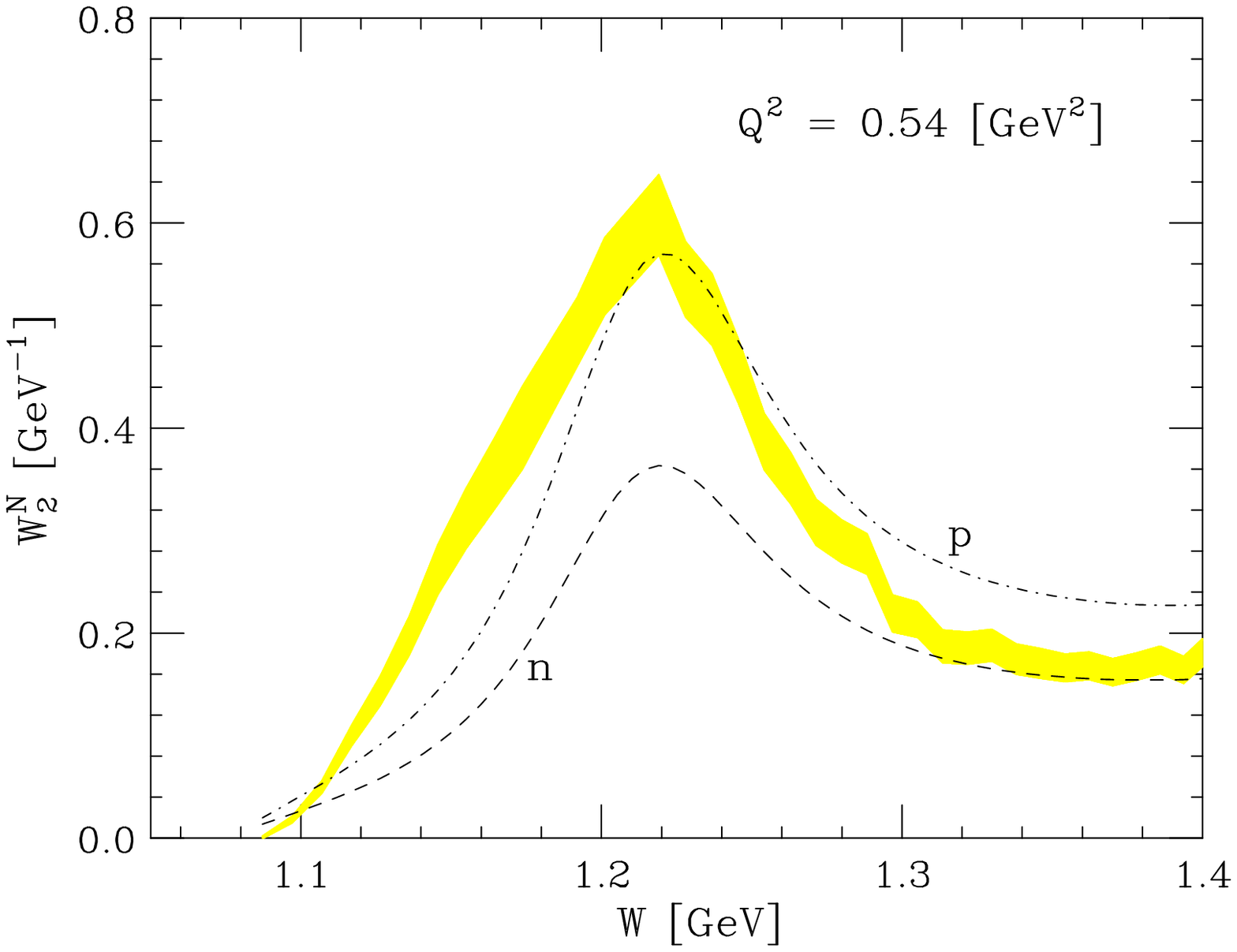,angle=00,width=6.2cm,height=5.0cm}}
{\psfig{figure=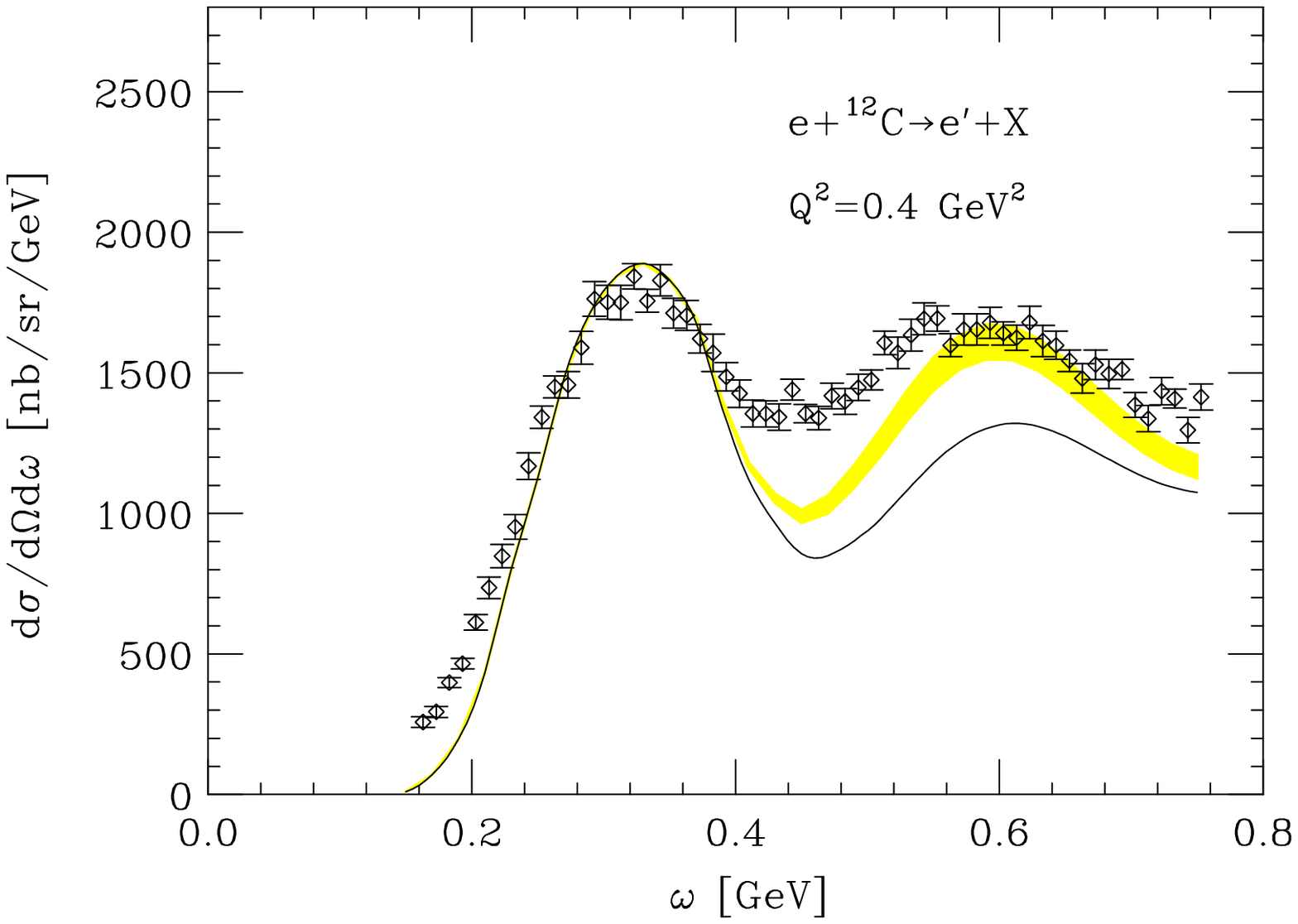,angle=00,width=6.2cm,height=5.0cm}}
\vspace*{8pt}
\caption{Left panel: nucleon structure functions $w^N_2$ ($N=n,p$) at $E_e = 2.445$ GeV 
and $\theta_e = 20^\circ$,
corresponding to $Q^2=0.54$ GeV$^2$ at the $\Delta$ production peak, plotted as
a function of the invariant mass of the hadronic final state. The shaded area
represents the $w^N_2$ resulting from the analysis of Ref.\protect\cite{bm1}, 
while the dashed and dot-dash lines correspond to $w^n_2$ and $w^p_2$ of 
Ref.\protect\cite{BR}, respectively.  
Right panel: electron scattering cross section off carbon at
$E_e = 1.3$ GeV and $\theta_e = 37.5^\circ$, as a function of the electron energy
loss $\omega$. The solid line corresponds to theoretical calculations carried out
using the proton and neutron structure functions of Ref. \protect\cite{BR}, while
the shaded region has been obtained using $w_1^n$ and $w_2^n$ of Ref. \protect\cite{bm1}.
The data are taken from Ref. \protect\cite{Sealock} \label{resonance} }
\end{figure}

The left panel of Fig. \ref{resonance} shows that the neutron structure function
extracted from Jefferson Lab data at $Q^2 \sim 0.5$ GeV$^2$\cite{JLab}, 
following the procedure of Bodek and Ritchie, 
is significatly larger than the one resulting from the 
analysis of Ref.\cite{BR}, based on SLAC data spanning the 
kinematical domain $1~<~Q^2~<~20$~GeV$^2$. 
The theoretical cross section obtained using the neutron structure function 
of Ref.\cite{bm1}, displayed in the right panel, turns out to be in close agreement with 
the SLAC data of Ref.\cite{Sealock}, corresponding to $Q^2 \sim 0.4$, in the $\Delta$ region.

\section{Charged current neutrino-nucleus interactions}
\label{nuA}

In Born approximation, the cross section of the weak charged current process
\beq
\nu_\ell + A \rightarrow \ell^- + X\ ,
\label{nu:process}
\eeq
can be written in the form (compare to Eq. (\ref{eA:xsec}))
\beq
\frac{d\sigma}{d\Omega_\ell dE_\ell} = \frac{G^2}{32 \pi^2}\
\frac{|{\bf k}^\prime|}{|{\bf k}|}\
 L_{\mu \nu} W^{\mu \nu}\ ,
\label{nu:cross:section}
\eeq
where $G=G_F \cos \theta_C$, $G_F$ and $\theta_C$ being Fermi's coupling constant and
Cabibbo's angle, $E_\ell$ is the energy of the final state
lepton and ${\bf k}$ and ${\bf k}^\prime$ are the neutrino and charged lepton
momenta, respectively. Compared to the corresponding quantities appearing in
Eq. (\ref{eA:xsec}), the tensors $L_{\mu \nu}$ and
$W^{\mu \nu}$ include additional terms resulting from the
presence of axial-vector components in the leptonic and hadronic
currents (see, e.g., Ref. \cite{walecka}).

Within the IA scheme, the cross section of Eq. (\ref{nu:cross:section}) can be cast
in a form similar to that obtained for the case of electron-nucleus scattering. 
Hence, its calculation requires the nuclear spectral function
and the tensor describing the weak charged current interaction of a free nucleon,
$w_N^{\mu\nu}$. In the case of quasi-elastic scattering, neglecting the contribution
associated with the pseudoscalar form factor $F_P$, the latter can be written
in terms of the nucleon Dirac and Pauli form factors $F_1$ and $F_2$, related to the
measured electric and magnetic form factors $G_{E}$ and $G_{M}$ through
\beq
F_{1} = \frac{1}{1 - q^2/4 m^2} \left( G_{E} - \frac{q^2}{4 m^2} G_{M} \right) \ \ , \ \ 
F_{2} = \frac{1}{1 - q^2/4 m^2} \left( G_{M} - G_{E} \right) \ ,
\eeq
and the axial form factor $F_A$.

It has to be pointed out that the formalism described in
Section \ref{SQW}, while including dynamical
correlations in the final state, does not take into account statistical
correlations, leading to Pauli blocking of the phase space available to the
knocked-out nucleon.

A rather crude prescription to estimate the effect of Pauli blocking amounts to
modifying the spectral function through the replacement
\beq
P({\bf p},E) \rightarrow P({\bf p},E)
\theta(|{\bf p} + {\bf q}| - {\overline p}_F)
\label{pauli}
\eeq
where ${\overline p}_F$ is the average nuclear Fermi momentum, defined as
\beq
{\overline p}_F = \int  d^3r\ \rho_A({\bf r}) p_F({\bf r}),
\label{local:kF}
\eeq
with $p_F({\bf r})=(3 \pi^2 \rho_A({\bf r})/2 )^{1/3}$, $\rho_A({\bf r})$ being the
nuclear density distribution. For oxygen, Eq. (\ref{local:kF}) yields
${\overline p}_F = 209$ MeV. Note that, unlike the spectral function, the
quantity defined in Eq. (\ref{pauli})
does not describe intrinsic properties of the target only, as it depends
explicitely on momentum transfer.

The effect of Pauli blocking is hardly visible in the energy loss spectra
shown in Fig. \ref{fig:ee}, as the kinematical setup corresponds to
$Q^2 > 0.2$ GeV$^2$ at the quasi-elastic peak. 
However, it becomes appreciable at lower $Q^2$.

\begin{figure}[hbt]
{\psfig{figure=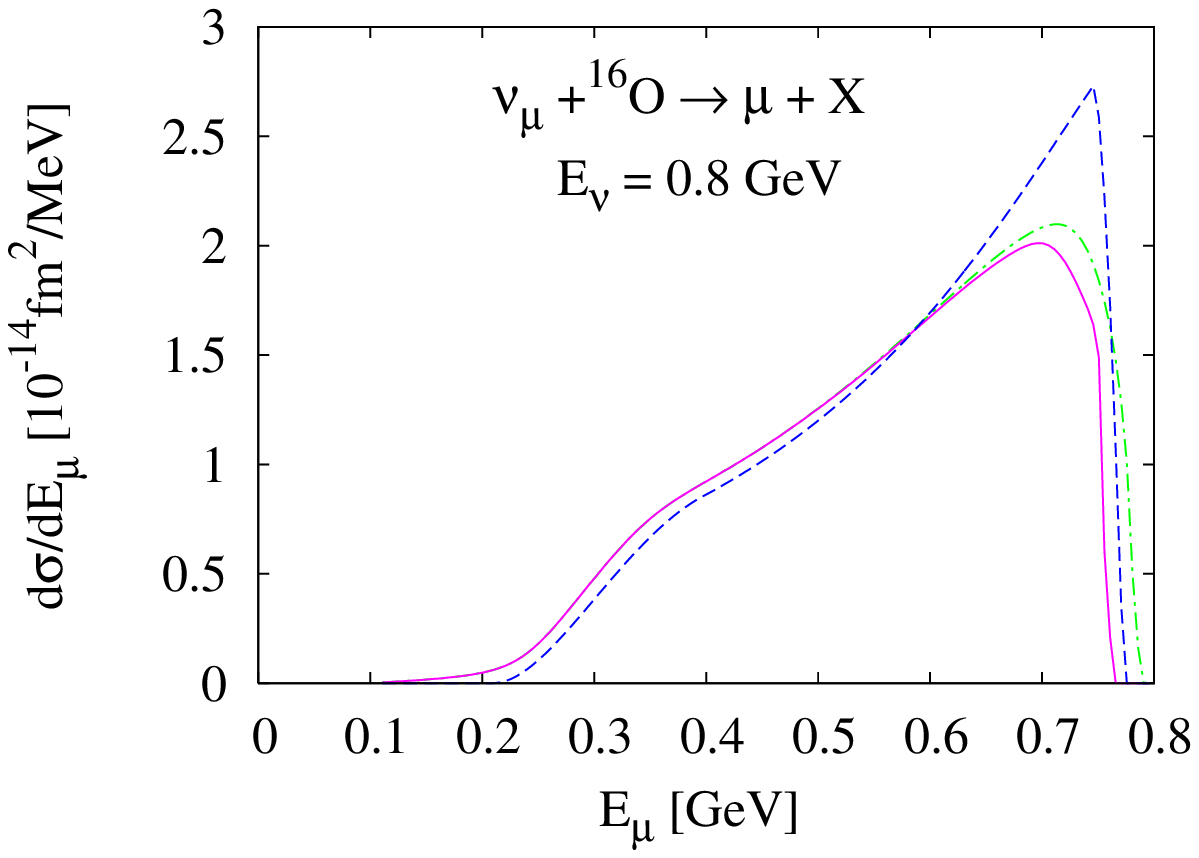,angle=00,width=6.2cm,height=5.0cm}}
{\psfig{figure=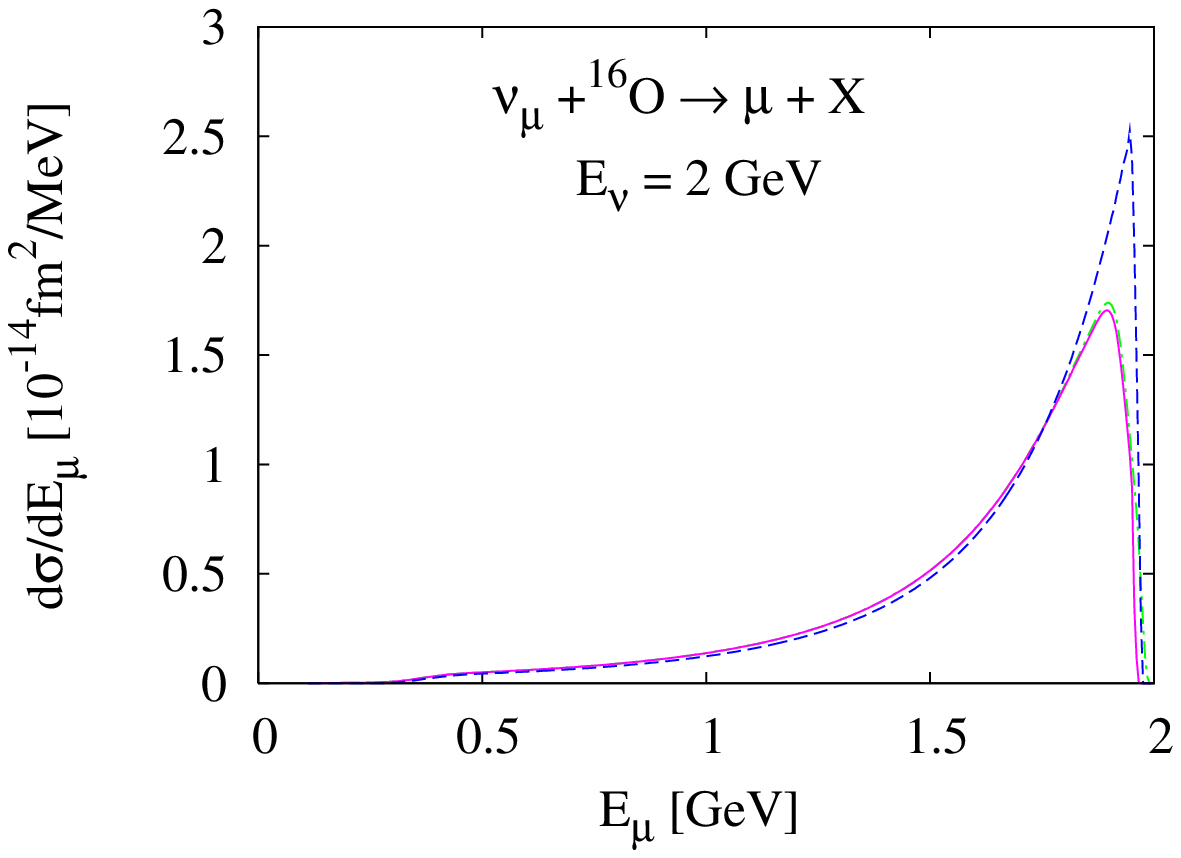,angle=00,width=6.2cm,height=5.0cm}}
\vspace*{8pt}
\caption{ Quasi-elastic differential cross section $d\sigma/dE_\mu$, 
as a function of the charged lepton energy $E_\mu$, for neutrino energy $E =$ 0.8
(left panel) and 2.0 GeV (right panel). The solid and dot-dash lines show the results of the 
IA calculation with and without Pauli blocking (implemented as in Eq. (\protect\ref{pauli})),
respsctively. The dashed lines have been obtained using the FG model. \label{del}}
\end{figure}

As an example, Fig. \ref{del} shows the $\nu_\mu$-nucleus cross sections,
as a function of the scattered muon energy, calculated within the FG model (dashed line) 
and using the the spectral functio of Ref.\cite{BFFS}, with and without Pauli blocking
(solid and dot-dash lines, respectively).  It clearly appears 
that the FG model yields a larger peak at high-energy. This feature should show
up in the cross section at forward angles, and  may have a direct
effect on neutrino oscillation measurements.

\section{Summary}
\label{summa}

The approach based on NMBT provides a unified parameter-free description of
the electroweak nuclear response in a variety of kinematical regions.

Thanks to the availability of reliable spectral functions, accurate calculations
of the cross sections in the IA regime can be carried including the
effects of short range NN correlations.
Correlation effects can also be consistenly included in the treatment of FSI
between the struck nucleon and the spectator particles.

Comparison to electron-nucleus scattering data shows that,
the region of quasi-elastic scattering and $\Delta$ production can 
be reproduced with remarkable accuracy. 
The role of meson exchange currents, which are known to provide a significant
amount of strength in the dip region between the quasi-elastic and the
$\Delta$ peak, still needs to be carefully investigated.
The generalization of the formalism discussed in these lectures to 
deep inelastic scattering is straightforward.\cite{BDS}

As a final remark, it has to be pointed out that the possibility of
using the approach based on NMBT in the analysis of neutrino experiments largely
depends on the ability to implement its elements in Monte Carlo simulations.

Assuming, for the sake of simplicity, that the elementary weak
interaction vertex in the nuclear medium be the same as in free space,
a realistic simulation of neutrino-nucleus scattering
requires the energy and momentum probability distribution of
the nucleons, needed to specify the initial state, as well as their
distribution in space and the medium modified hadronic cross section, needed
for the description of FSI.

Studies based on NMBT and stochastic methods to solve the many-body
Sch\-r\"o\-din\-ger equation appear to be capable of providing access to all
the above quantities for a variety of nuclear targets.

\section*{Acknowledgements}

These lectures are dedicated to the memory of Adelchi Fabrocini and Vijay Pandharipande,
whose work led to important and lasting progress in the many-body theory of the 
nuclear response.



\begin{thebibliography}{0}

\bibitem{BDS}
O. Benhar, D. Day and I. Sick, {\it Rev. Mod. Phys.}, {\bf 80} (2008) 189.

\bibitem{eepreviews}
Proceedings of the Fifth International Conference on Perspectives in Hadronic Physics, 
Edited by C. Ciofi degli Atti and D. Treleani, {\it Nucl. Phys. A} {\bf 782} (2007).

\bibitem{nuint}
Proceedings of The Fifth International Workshop on Neutrino-Nucleus
Interactions in the Few-GeV Region (NuInt07), 
Edited by G.P. Zeller, J.G. Morfin and F. Cavanna (AIP, New York, 2007).

\bibitem{WSS}
R.B.~Wiringa , V.G.J.~Stoks, R.~Schiavilla,  
{\it Phys. Rev.} {\bf C51} (1995) 38.

\bibitem{PPCPW}
P.S.~Pudliner {\it et al}, 
{\it Phys. Rev.}  {\bf C56} (1997) 1720.

\bibitem{WP}
S.C.~Pieper and R.B.~Wiringa, 
{\it Ann. Rev. Nucl. Part. Sci.} {\bf 51} (2001) 53.

\bibitem{DayRev}
B.D.~Day, {\it Rev. Mod. Phys.} {\bf 39} (1967) 719;
{\it ibidem} {\bf 50} (1978) 495.

\bibitem{Fee69}
E.~Feenberg, {\it Theory of Quantum Fluids}, (Academic Press, New York, 1969).

\bibitem{CMK79}
J.W.~Clark, {\it Prog. Part. Nucl. Phys.} {\bf 2} (1979) 89.

\bibitem{FPa88}
S.~Fantoni and V.R.~Pandharipande, {\it Phys. Rev. C} {\bf 37} (1988) 1697.

\bibitem{FetterWalecka}
A.~Fetter, and J.~Walecka, {\it Quantum Theory of Many Particle Systems},
  (McGraw-Hill, New York, 1971).

\bibitem{BFF:1992}
O.~Benhar, A.~Fabrocini, and S.~Fantoni, {\it Nucl. Phys. A} {\bf 550} (1992) 201.

\bibitem{BFF:1989}
O.~Benhar, A.~Fabrocini, and S.~Fantoni, {\it Nucl. Phys. A} {\bf 505} (1989) 267.

\bibitem{BFF:1990}
O.~Benhar, A.~Fabrocini, and S.~Fantoni, {\it Phys. Rev. C} {\bf 41} (1990) R24.

\bibitem{WIM:2004}
W.~Dickhoff, and C.~Barbieri, {\it Prog. Part. Nucl. Phys.} {\bf 52} (2004) 377.

\bibitem{gangofsix}
O.~Benhar {\it et al}, 
{\it Phys. Rev. C} {\bf 44} (1991) 2328.

\bibitem{marianthi}
M.~Petraki {\it et al},
{\it Phys. Rev. C} {\bf 67} (2001) 014605.

\bibitem{papi}
V.R.~Pandharipande and S.C.~Pieper, {\it Phys. Rev. C} {\bf 45} (1992) 791.

\bibitem{Itzykson80}
C.~Itzykson and J.~Zuber, {\it Quantum Field Theory} (McGraw-Hill, New York, 1980).

\bibitem{Golak95}
J~Golak {\em et~al.}, {\it Phys. Rev. C} {\bf 52} (1995) 1216.

\bibitem{Efros94}
V.D.~Efros, W.~Leidemann, and G.~Orlandini, {\it Phys. Lett.} {\bf B338} (1994) 130.

\bibitem{Carlson98}
J.~Carlson and R.~Schiavilla, {\it Rev. Mod. Phys.} {\bf 70} (1998 ) 743.

\bibitem{Leidemann}
W. Leidemann, these Proceedings.

\bibitem{FP}
S.~Fantoni, and V.R.~Pandharipande, Nucl. Phys. {\bf A473} (1987) 234.

\bibitem{FF}
A.~Fabrocini and S.~Fantoni, Nucl. Phys. {\bf A503} (1989) 375.

\bibitem{Dellafiore85}
A.~Dellafiore, F.~Lenz and F~Brieva, {\it Phys. Rev. C} {\bf 31} (1985) 1088.

\bibitem{BR}
A.~Bodek and J.~ Ritchie, {\it Phys. Rev. D} {\bf 23} (1981) 1070.

\bibitem{bm1}
O.~Benhar and D.~Meloni, {\it Phys. Rev. Lett.} {\bf 97} (2006) 192301.

\bibitem{bm2}
O.~Benhar and D.~Meloni, {\it Nucl. Phys. A} {\bf 789} (2007) 379.

\bibitem{LNF}
M.~Anghinolfi {\it et al.}, {\it Nucl. Phys. A} {\bf 602} (1996) 405.

\bibitem{Benhar05}
O.~Benhar {\it et al}, 
{\it Phys. Rev. D} {\bf 72} (2005) 053005.

\bibitem{BFFS}
O.~Benhar, A.~Fabrocini, S.~Fantoni and I.~Sick, {\it Nucl. Phys. A} 
{\bf 579} (1994) 493.

\bibitem{Hohler76} 
G.~H\"ohler {\it et al.} {\it Nucl.\ Phys. B} {\bf 114} (1976) 505.

\bibitem{Brash02} E.J.~Brash, A.~Kozlov, Sh.~Li, and G.M.~Huber,
{\it Phys. Rev. C} {\bf 65} (2002) 051001(R).

\bibitem{hiroki}
H.~ Nakamura, M.~Sakuda, T.~Nasu and O.~Benhar, 
{\it Phys. Rev. C} {\bf 76} (2007) 065208.

\bibitem{JLab}
I. Niculescu {\it et al}, {\it Phys. Rev. Lett.} {\bf 85} (2000) 1186.

\bibitem{Sealock}
R.M. Sealock {\it et al}, {\it Phys. Rev. Lett.} {\bf 62} (1989) 1350.

\bibitem{walecka} 
J.D. Walecka, {\it The\-o\-re\-ti\-cal Nu\-cle\-ar and Sub\-nu\-cle\-ar
Physics} (Oxford University Press, 1995).








\end{thebibliography}
\end{document}